\documentclass[a4paper,11pt]{article}
\usepackage{pos}

\newcommand{\mean}[1]{\left\langle{#1}\right\rangle}
\newcommand{\condl}{\mean{\bar q q}_l}
\newcommand{\conds}{\langle \bar s s \rangle}

\title{Chiral Symmetry Restoration, Thermal Resonances and the $U(1)_A$ symmetry}

\author*{Angel G\'omez Nicola}
\author{Jacobo Ruiz de Elvira}
\author{Andrea Vioque-Rodr\'iguez}

\affiliation{Universidad Complutense de Madrid, Facultad de Ciencias F\'isicas, Departamento de F\'isica Te\'orica  and IPARCOS,
  Plaza de las Ciencias 1, 28040 Madrid, Spain}

\emailAdd{gomez@ucm.es}

\abstract{We discuss recent results regarding the interplay between chiral and $U(1)_A$ symmetry restoration, both from the point of view of Ward Identities relating meson susceptibilities and quark condensates, and from the behaviour of light meson resonances at finite temperature.}

\FullConference{%
  *** Particles and Nuclei International Conference - PANIC2021 ***\\
  *** 5 - 10 September, 2021 ***\\
  *** Online ***
}

\begin{document}
\maketitle

\section{Introduction}
Restoration of the $SU(N_f)_V\times SU(N_f)_A$ chiral symmetry of the massless QCD lagrangian, with $N_f$ light flavours, has become a central problem for the study of the QCD phase diagram \cite{Ratti:2018ksb,Nicola:2020smo}. Lattice analyses \cite{Aoki:2009sc,Bazavov:2011nk}  support a crossover transition in the physical limit at finite temperature $T_c\sim$ 155 MeV and vanishing baryon chemical potential, which  presumably degenerates into a second-order phase transition for $N_f=2$ massless quarks (light chiral limit) \cite{Pisarski:1983ms}. An still open problem, though, with important physical consequences  regarding the nature of the transition, is the connection  with the asymptotic restoration of the anomalous $U(1)_A$ symmetry \cite{Pisarski:1983ms,Shuryak:1993ee,Pelissetto:2013hqa}. Thus, while lattice analyses of $N_f=2+1$  susceptibilities near the physical case show a sizable temperature gap between the degeneration of chiral partners such as $\pi-\sigma$ and that of $U(1)_A$ partners such as $\pi-\delta$ \cite{Buchoff:2013nra}, $N_f=2$ studies  \cite{Brandt:2016daq} for screening masses are compatible with $U(1)_A$ restoration at the chiral transition in the light chiral limit, showing  a much smaller gap  for physical masses.  Here, $\pi,\sigma,\delta$ stand for the quark bilinears corresponding to the pion, the light component of $f_0(500)/f_0(980)$  pair and the $a_0(980)$ respectively.  In the present work we will review recent theoretical advances aiming to reconcile those lattice results, using Ward Identities and thermal resonances.  

\section{Ward Identities}
Ward Identities (WI) derived from the QCD generating functional can be used to relate susceptibilities and quark condensates in different channels 
\cite{Nicola:2013vma,Nicola:2016jlj,GomezNicola:2017bhm,Nicola:2018vug}  which is quite useful for the problem at hand. A particularly relevant example is the following relation between pseudoscalar susceptibilites,
\begin{equation}
\chi_P^{ls}(T)=- \frac{1}{2}\frac{m_l}{m_s}  \left[\chi_P^\pi (T)-\chi_P^{ll}(T)\right]\equiv -2\frac{m_l}{m_s} \chi_{5,disc} ,
\label{WI1}
\end{equation}
where $\chi_P^{\pi,ll,ls}$ stand respectively for the susceptibilities of the $\pi\pi,\eta_l\eta_l,\eta_l\eta_s$ correlators, with $\eta_{l,s}$ the light and strange components of the $\eta$ state.  It is possible to choose a $SU(2)_A$ rotation transforming the pseudoscalar $\eta_l$ into the scalar $\delta$ (those states are chiral partners) and therefore if the chiral symmetry was effectively restored, the $ls$ correlator should vanish by a parity-conserving argument. But then, according to \eqref{WI1}, $\chi_P^\pi \rightarrow \chi_P^{ll}$ in that regime and those $U(1)_A$ partners would degenerate as well. This argument leads then to exact $U(1)_A$ restoration whenever chiral symmetry is also exactly restored, as for two massless flavours, consistently with the lattice analyses previously mentioned.

Other identities specially useful are those in the $I=1/2$ $K-\kappa$ sector, whose lightest meson states are the kaon and the $K_0^* (700)$, namely, 
\begin{eqnarray}
\chi_P^K(T)=-\frac{\condl (T)+2\conds (T)}{m_l + m_s}, \quad 
\chi_S^\kappa (T)=\frac{\condl (T)-2\conds (T)}{m_s-m_l},
  \label{wikappa} 
\end{eqnarray}
 
 The light quark condensate $\condl$ decreases rapidly with temperature around $T_c$, where it develops an inflection point in the physical limit, while the strange condensate $\conds$ remains almost constant up to $T_c$ and decreases slowly beyond that temperature \cite{Bazavov:2011nk}. From that behaviour and the identities \eqref{wikappa} one infers that   the susceptibility in the $\kappa$ channel should have a peak above  the chiral transition, beyond which it decreases, becoming degenerate with the kaon susceptibility. Such degeneration signals also $O(4)\times U(1)_A$ restoration and allows to understand the role of strangeness through $\conds$ in \eqref{wikappa} \cite{GomezNicola:2020qxo}. 
 
\begin{figure}
  \centerline{
    \includegraphics[width=0.45\textwidth]{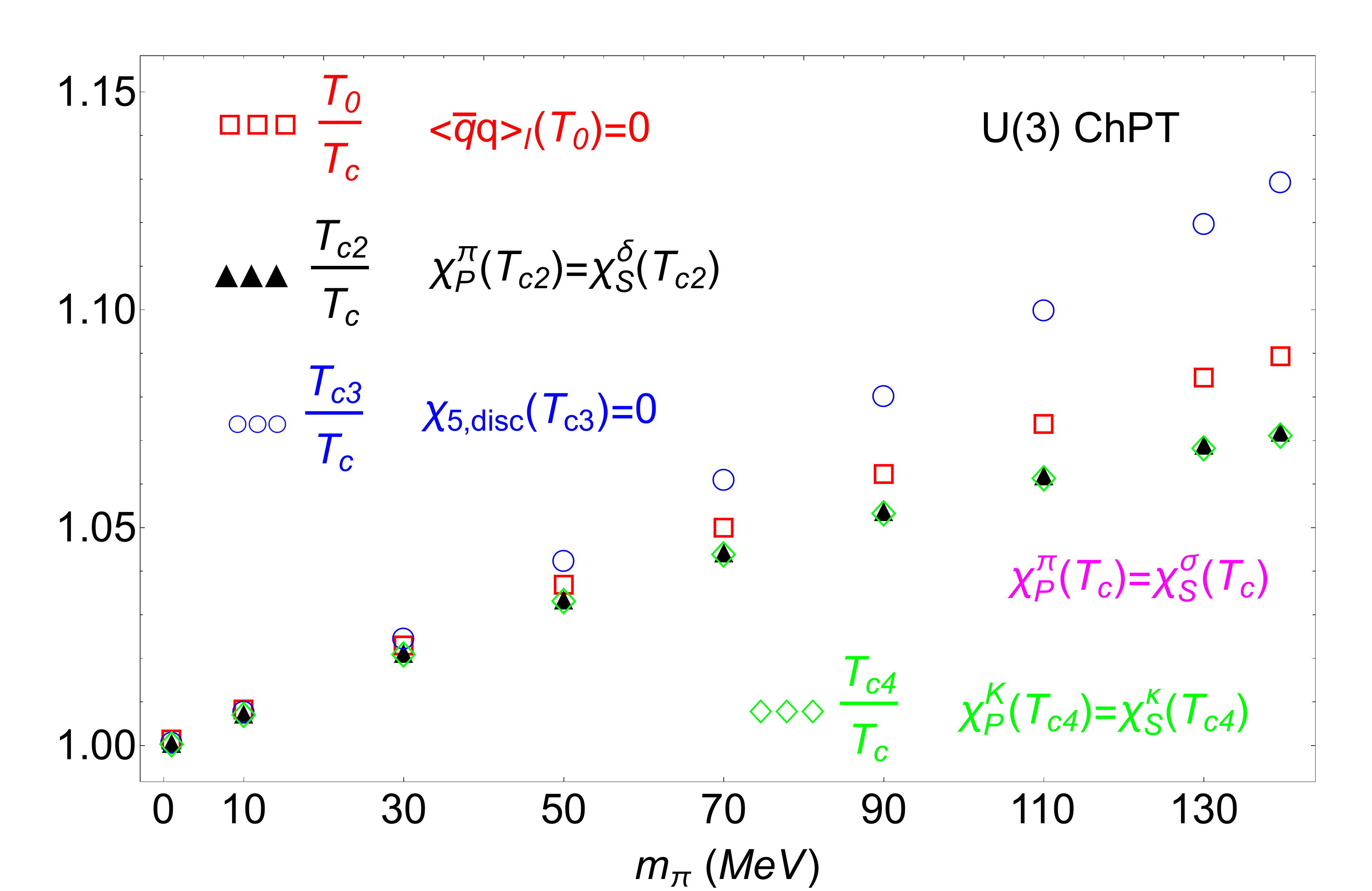}
    \includegraphics[width=0.45\textwidth]{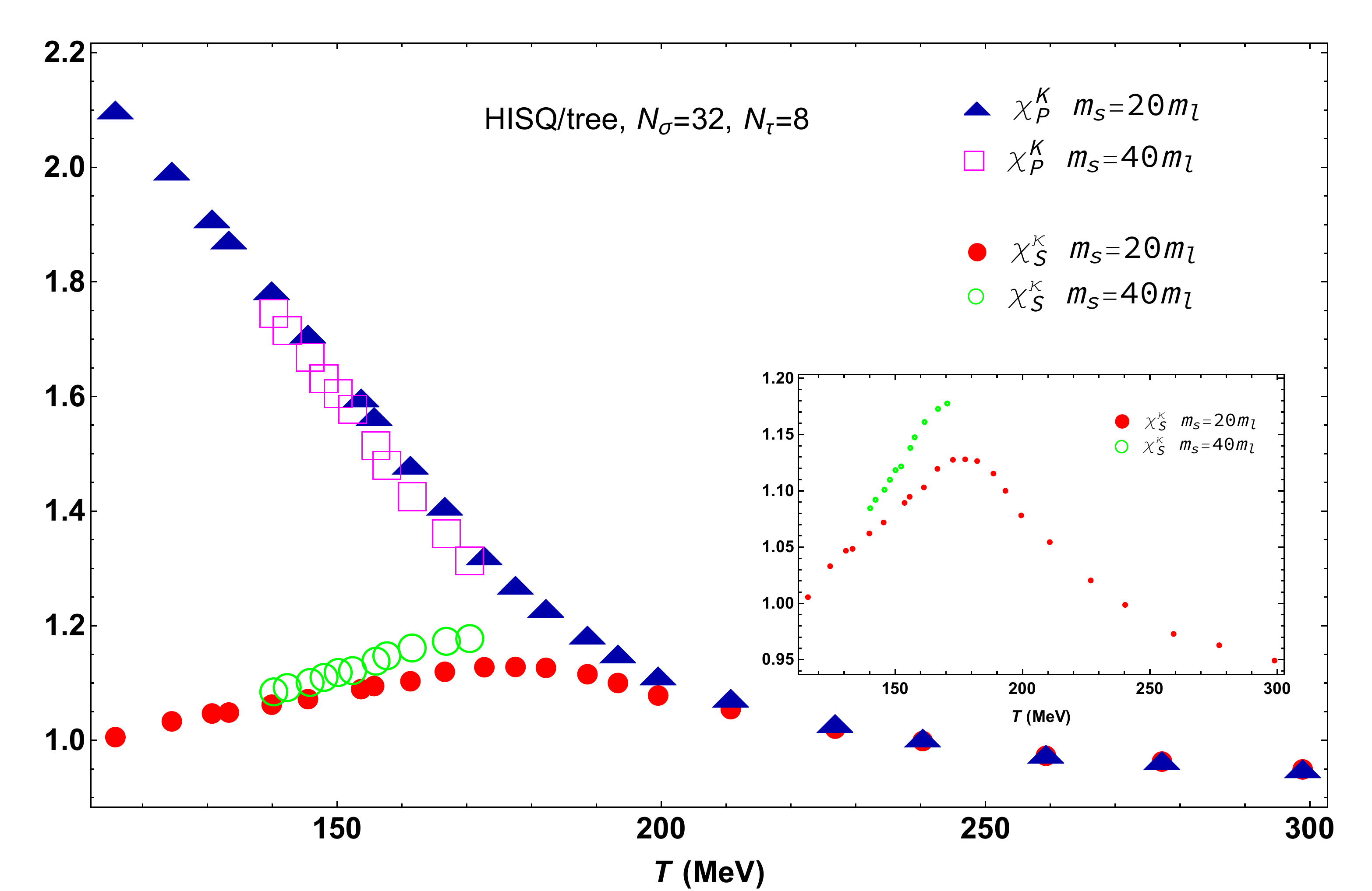}
   } 
  \caption{Left: Results from the $U(3)$ ChPT analysis in  \cite{Nicola:2018vug}  for the scalar/pseudoscalar nonet susceptibilities. Right:  reconstructed pseudoscalar and scalar susceptibilities \cite{GomezNicola:2020qxo} in the $I=1/2$ channel from \eqref{wikappa}, with 
    light- and strange-quark condensate data from~\cite{Bazavov:2011nk}, in lattice units.}
   \label{fig:chis}
\end{figure}

 To illustrate the above results, we show in Figure \ref{fig:chis}, on the one hand, the behaviour of  pseudo-critical degeneration temperatures for different $O(4)$ and $O(4)\times U(1)_A$ partners, calculated within the $U(3)$ Chiral Perturbation Theory (ChPT) framework \cite{Nicola:2018vug}, which confirms that all those temperatures approach the same value as the light chiral limit is approached. On the other hand, we show the result in \cite{GomezNicola:2020qxo} 
  for the reconstructed $K,\kappa$ susceptibilities using the WIs in \eqref{wikappa} and lattice  condensates, since there are no direct susceptibility data available in that channel.  The peak behaviour for $\chi_S^\kappa$ and the degeneration pattern $\chi_S^\kappa-\chi_P^K$ are clearly observed. 
\section{Thermal resonances}
The previously discussed trends have been also analyzed from the behaviour of resonances generated dynamically from Unitarized ChPT at finite temperature. The connection with scalar susceptibilities has been carried out saturating those susceptibilities with the lightest $f_0(500)$ and $K_0^* (700)$ poles in the second Riemann sheet of the elastic $\pi\pi$ and $\pi K$ scattering respectively, unitarized at finite temperature  \cite{Nicola:2013vma,Ferreres-Sole:2018djq,GomezNicola:2020qxo}.

\begin{figure}[h]
  \centerline{
    \includegraphics[width=0.45\textwidth]{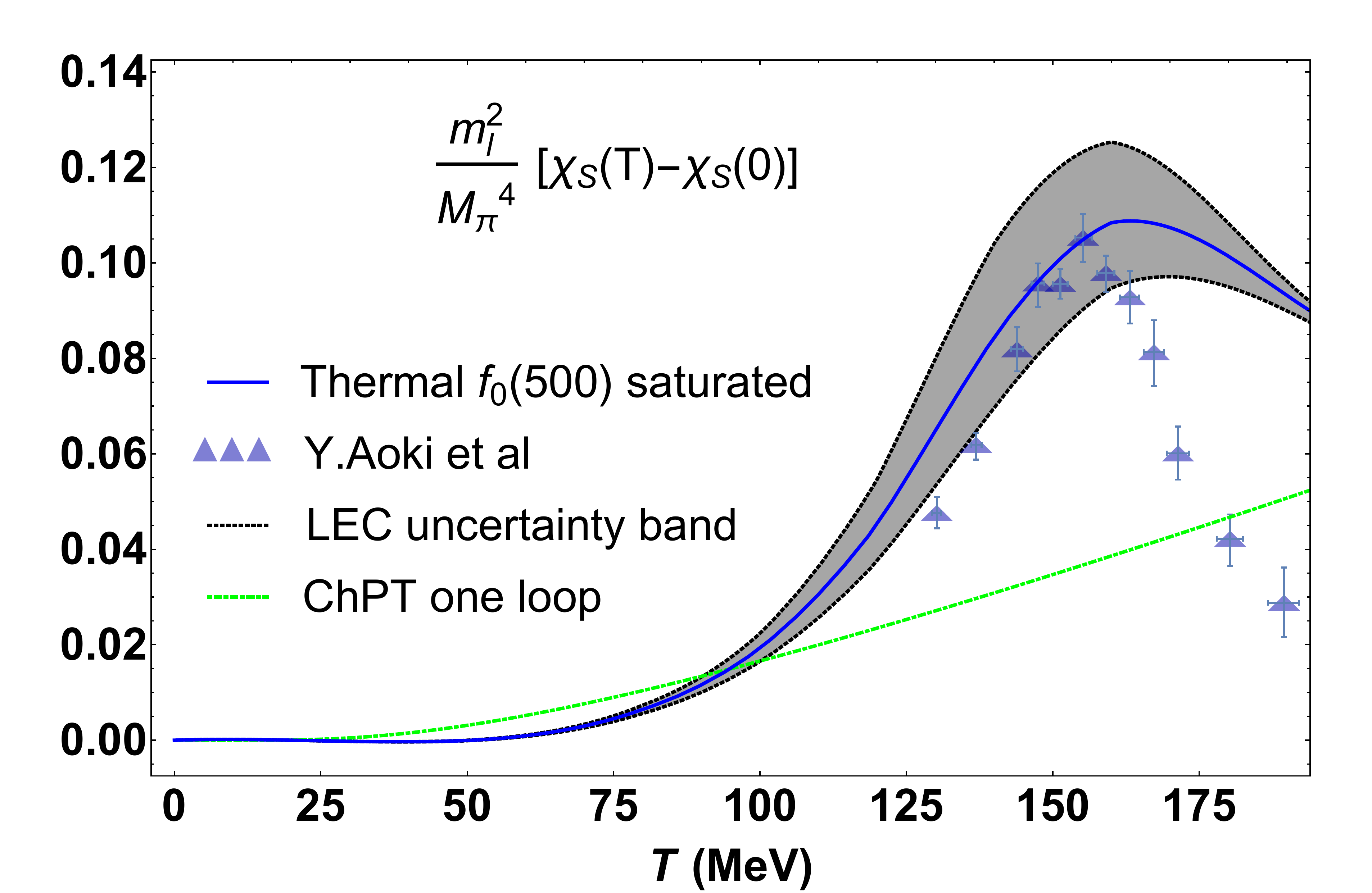}
    \includegraphics[width=0.45\textwidth]{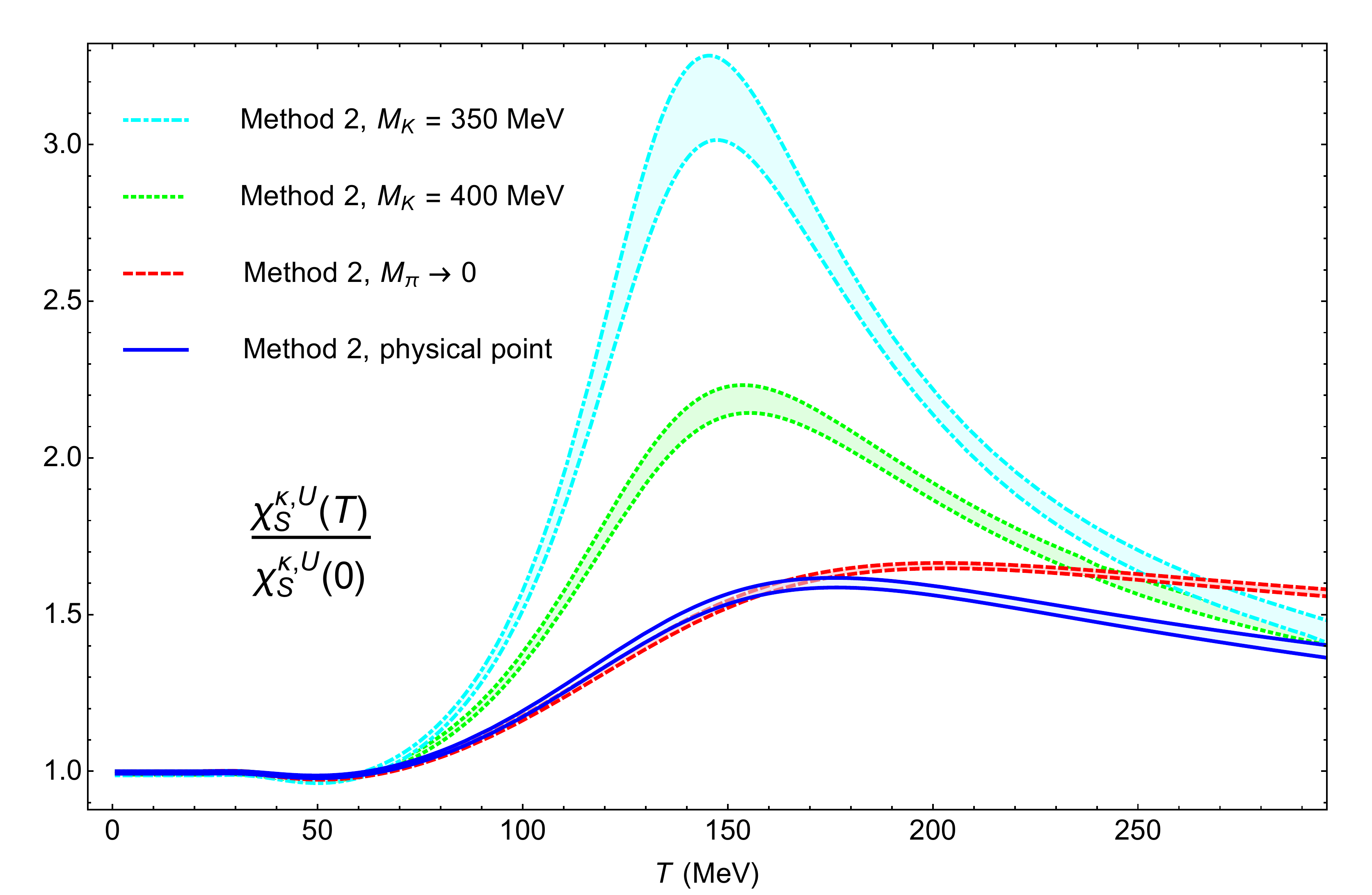}
    }
\caption{Scalar susceptibilities saturated by thermal resonances in the $I=0$ (left) and $I=1/2$ (right) channels. Lattice points in the left pannel are taken from \cite{Aoki:2009sc}.} 
\label{fig:resonances}       
\end{figure}

In Figure \ref{fig:resonances} we show the result of such approach.  The $I=0$ ($\sigma$ channel) scalar susceptibility develops a peak around the transition, as expected, compatible with lattice points below and around $T_c$ within the uncertainties of the involved Low-Energy Constants (LEC). The $I=1/2$ susceptibility in the $\kappa$ channel shows also the expected peak above $T_c$,  flattened above the peak  as  $m_l/m_s$ is reduced, indicating a more rapid degeneration with the $K$ channel, and enhanced as $m_l/m_s$ tends to unity, which is  the $SU(3)$  limit where the $\kappa$ and $\sigma$ channels are expected to show a similar behaviour. 
\section{Conclusions}
Recent analyses within Ward Identities, ChPT and thermal resonances yield consistent results regarding chiral and $U(1)_A$ restoration at finite temperature. Exact chiral restoration would lead to degeneration of $U(1)_A$ partners, while the behaviour of the $\kappa$ scalar susceptibility reveals the role of strangeness in the physical mass case, reconciling lattice results for two and three light flavours. 

\begin{acknowledgments}
 Work partially supported by research contract  PID2019-106080GB-C21 ( ``Ministerio de Ciencia e Innovaci\'on"),  the European Union Horizon 2020 research and innovation program under grant agreement No 824093 and the Swiss National Science Foundation, project No.\ PZ00P2\_174228. A. V-R acknowledges support from a fellowship of the UCM predoctoral program.
 \end{acknowledgments}

\end{document}